\documentclass[slac_one]{revtex4}
\usepackage{graphicx}
\usepackage{fancyhdr}
\renewcommand{\headrulewidth}{0pt}
\renewcommand{\footrulewidth}{0pt}

\newcommand{\EE}{\rm e^+ e^-}

\begin{document}

%Title of paper
\title{{\small{2005 International Linear Collider Workshop - Stanford,
U.S.A.}}\\ %% Please keep this conference title here
\vspace{12pt}
Two-photon width of the Higgs boson} %% Paper title goes here

% Repeat the \author .. \affiliation  etc. as needed
%
% \affiliation command applies to all authors since the last
% \affiliation command. The \affiliation command should follow the
% other information

\author{K. M\"onig}
\affiliation{DESY, Zeuthen, D 15738, Germany}
\author{A. Rosca}
\affiliation{West University of Timisoara, Timisoara, RO 300223, Romania}

\begin{abstract}
This study investigates the potential of a photon collider
for measuring the two photon partial width times the
branching ratio of a light Higgs boson. The analysis is
based on the reconstruction of the Higgs events produced
in the $\gamma \gamma \to$h process, followed by Higgs
decay into a b${\rm \bar{b}}$ pair.
A statistical error of the measurement of the two-photon width times
the b${\rm \bar{b}}$ branching ratio of the Higgs boson is found to be 1.7 $\%$
with an integrated luminosity of 80 fb$^{-1}$ in the high energy part
of the spectrum. 
\end{abstract}

%\maketitle must follow title, authors, abstract
\maketitle

\thispagestyle{fancy}

% body of paper here - Use proper section commands
% References should be done using the \cite, \ref, and \label commands
% Put \label in argument of \section for cross-referencing
%\section{\label{}}

\section{Introduction} % Section title should be in all capitals.
The central challenge for particle physics nowadays is the origin
of mass. In the Standard Model both fermions and gauge boson masses
are generated through interactions with the same scalar particle,
the Higgs boson, h. If it exists, the Higgs boson will certainly
be discovered by the time a photon collider is constructed. The aim of
this machine will be then a precise measurement of the Higgs 
properties. The photon scattering can be used to produce the Higgs 
particles singly in the s-channel of the colliding photons.
This facility permits a high precision measurement of the 
h $\to \gamma \gamma$ partial width, which is sensitive to new charged
particles.
The measurement is significantly important. If we find a deviation of
the two photon width from the standard model prediction it means  
that an additional contribution from unknown particles is present, and 
so it is a signature of physics beyond the SM. For example,
the minimal extension of the SM predicts the ratio of the two photon width
$\Gamma(\rm h \to \gamma \gamma, \rm MSSM)$/
$\Gamma(\rm h \to \gamma \gamma, \rm SM) < $ 1.2  \cite{mssm}
for a Higgs boson with a mass of 120 GeV, assuming a supersymmetry  
scale of 1 TeV and the chargino
mass parameters $M$ and $\mu$ of 300 and 100 GeV, respectively.

A photon collider can measure the product 
$\Gamma(\rm h \to \gamma \gamma)$$\times$BR$(\rm h \to \rm X)$. To
obtain the two-photon partial width independent of the
branching ratio one has to combine with an accurate
measurement of the BR($\rm h \to \rm X$) from another
machine.

This study investigates the accuracy of the measurement of the two
photon decay width times the branching ratio 
for a Higgs boson with the mass of 120 GeV, the
preferred mass region by recent electroweak data \cite{ew}.
The signal and background processes studied are specified in section 2.
Event selection is described in section 3.
Results are summarized in section 4.

The feasibility of the measurement of the two photon decay width of the 
Higgs boson in
this mass region has also been reported by \cite{notes}.

\section{Simulation of the signal and background processes}

%\subsection{General Layout}

High energy photon beams can be produced at
a high rate in Compton backscattering of laser photons off
high energy electrons \cite{pc}. Setting opposite helicities for
the laser photons and the beam electrons the energy
spectrum of the backscattered photons is peaked at ~80$\%$
of the ${\rm e^{-}}$ beam energy. The backscattered photons 
are highly polarized in this high energy region.
With an integrated luminosity of 80 fb$^{-1}$ per year
for $\sqrt s_{ee} > 0.8 \sqrt {s_{ee}^{max}}$ \cite{pc}, about 20000
Higgs bosons with standard model coupling and a mass of 120 GeV 
can be produced in the $\gamma \gamma \to$h 
process. In this mass region the Higgs particle will
decay dominantly into a b${\rm \bar{b}}$ pair.

The beam spectra at $\sqrt s_{ee}$ = 210 GeV are simulated
using the CompAZ \cite{compaz}, a fast parameterization which includes 
multiple interactions and non-linearity effects.

The response of the detector has been simulated with
SIMDET 4 \cite{simdet}, a parametric Monte Carlo for the TESLA
$\EE$ detector. It includes a tracking and calorimeter simulation
and a reconstruction of energy-flow-objects (EFO). 
Only EFOs with a polar angle above 7$^{0}$ can be taken
for the Higgs reconstruction simulating the acceptance
of the photon collider detector as the only deference
to the $\EE$ detector \cite{ggdet}.

The considered backgrounds are the 
direct continuum $\gamma \gamma \to \rm b \bar{\rm b}$ and
$\gamma \gamma \to \rm c \bar{\rm c}$ production.
Due to helicity conservation, the
continuum background production proceeds mainly through states of 
opposite photon
helicities, making the states $J_{\rm z} = 2$. Choosing equal helicity 
photon
polarizations the
cross section of the continuum background is suppressed by a factor
$M_{\rm q}^{2}/s_{\gamma \gamma}$, with $M_{\rm q}$
being the quark mass. Unfortunately, this suppression does not apply
to the process $\gamma \gamma \to \rm q \bar{\rm q} \rm g$, because after 
the gluon
radiation the $\rm q \bar{\rm q}$ system is not necessarily in a
$J_{\rm z} = 0$
state. The surviving background is large and overwhelms the signal.

\begin{figure*}[t]
\centering
\includegraphics[width=120mm]{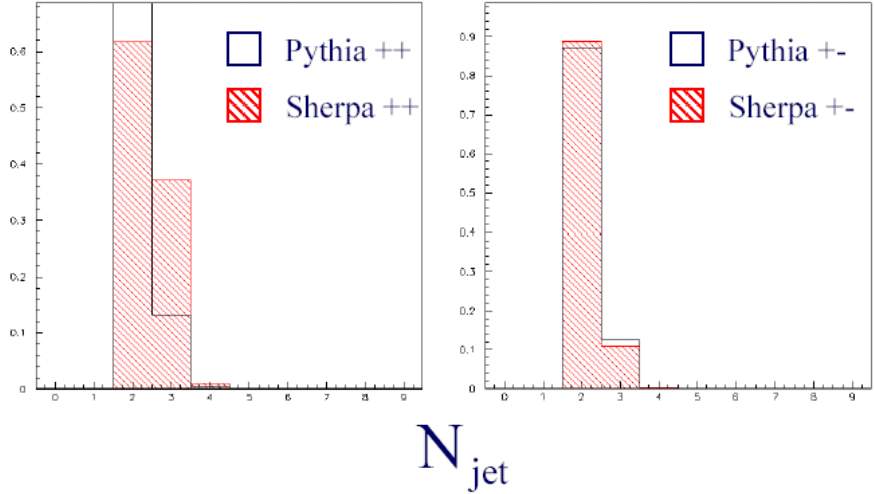}
\caption{Distributions for the number of jets simulated with PYTHIA and SHERPA
for the $\gamma \gamma \to \rm b \bar{\rm b} (\rm
g)$ and $\gamma \gamma \to \rm c \bar{\rm c} (\rm g)$ processes for $J_{z}$ = 0 and $J_{z}$ = 2.} 
\label{fig_jets}
\end{figure*}

\begin{figure*}[t]
\centering
\includegraphics[width=120mm]{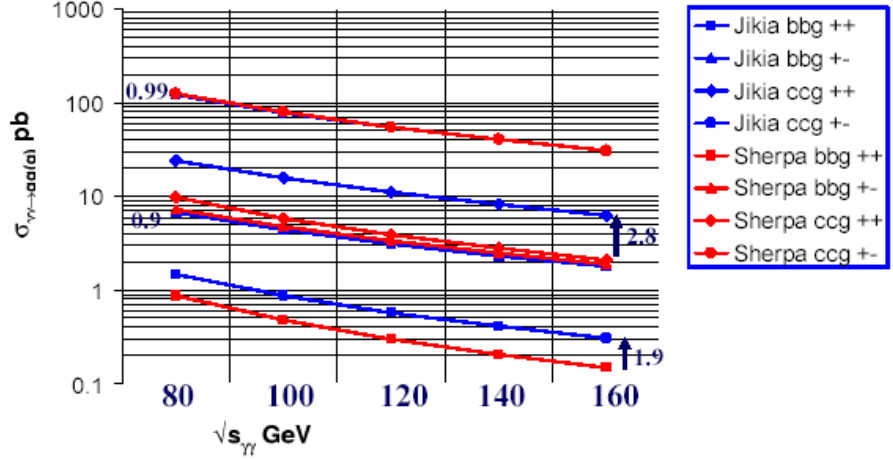}
\caption{Cross section comparison for the $\gamma \gamma \to \rm b \bar{\rm b} (\rm 
g)$ and $\gamma \gamma \to \rm c \bar{\rm c} (\rm g)$ processes as gives by SHERPA 
and the correct NLO values.} 
\label{fig_cs}
\end{figure*}

Signal $\gamma \gamma
\to {\rm h} \to {\rm b} \bar{\rm b}$ events are generated using
PYTHIA 6.2 \cite{pythia}. 
Background processes
$\gamma \gamma \to {\rm q} \bar{\rm q}$(g) are generated with the 
SHERPA (The simulation for High Energy Reactions of Particles) 
\cite{she} generator.
Higher order QCD effects are simulated in PYTHIA by evolving the
hard process event using the parton shower, which allows partons
to split into pairs of other partons. This technique is most
effective when the emitted gluons are soft or collinear, while
the region of high transverse momentum is poorly described.
For a reliable background estimation correct NLO corrections
are needed. Combining the hard process
with its higher order correction in PYTHIA is not trivial. A fraction
of $\gamma \gamma \to {\rm b} {\rm \bar{b}} {\rm g}$ events are included
in the $\gamma \gamma \to {\rm b} {\rm \bar{b}}$ process via gluon radiation 
in the parton shower. Combining the two processes without special 
procedures amounts in double counting of some phase space regions.
SHERPA \cite{she} is a generator which matches correctly the exact matrix
elements with showering.
A comparison between the distributions of jet multiplicity in the
$\gamma \gamma \to {\rm b} {\rm \bar{b}} {\rm (g)}$ process obtained
with PYTHIA and SHERPA is shown in Figure \ref{fig_jets}.
However, since virtual diagrams are not included, to obtain
the correct rates the SHERPA results for the spin $J_{z}=0$ states have to be scaled 
by the correct NLO cross section. The cross sections
at the NLO for the $\gamma \gamma \to {\rm q} {\rm \bar{q}} {\rm g}$
process have been calculated \cite{qcd1}, \cite{qcd2}. A cross section comparison 
is shown
in Figure \ref{fig_cs}.

\begin{table}[t]
\begin{center}
\caption{Cross sections and the number of expected and generated events 
for the signal and background processes}
\begin{tabular}{|l|c|}
\hline   & \textbf{Number of} \\
       &\textbf{selected events}   \\ \hline 
Signal process &
 \\
$\gamma \gamma \to \rm h \to {\rm b} \bar{\rm b}$
& 6044 \\
Background &    \\
$\gamma \gamma \to {\rm b} \bar{\rm b}$(g) & 1755.0 \\
$J$ = 0 &  \\
$\gamma \gamma \to {\rm b} \bar{\rm b}$(g)
 & 1865.0 \\
$J$ = 2 & \\
$\gamma \gamma \to {\rm c} \bar{\rm c}$(g) &298.6 \\
$J$ = 0 &  \\
$\gamma \gamma \to {\rm c} \bar{\rm c}$(g) &641.0\\
$J$ = 2  &\\
\hline
\end{tabular}
\label{tab_cs}
\end{center}
\end{table}

\section{Event selection}

The analysis aims to select events with two or three jets
from the Higgs boson decay. Two of these jets contain bottom quarks. The 
invariant
mass of the jets has to be consistent with the Higgs mass.

High multiplicity (FEO) events are selected and their visible
energy is required to be greater than 95 GeV. Events with
longitudinal imbalance greater than 10$\%$ of the visible energy
are rejected. 
Finally the cosine of the thrust angle has to be less than 0.7.

In the remaining event sample 
jets are reconstructed using the DURHAM clustering scheme \cite{durham}
with the resolution parameter y$_{\rm cut}$=0.02.

%Figures \ref{fig_costh}a,b shows the distribution of the cosine of the 
%thrust angle
%for the signal
%and background events. The s-channel signal process has an isotropic 
%angular distribution,
%while the t-channel background processes are forward peaked.

The cross section for the continuum production of the charm quark is 16 
times
larger than for bottom quarks, therefore b-quark tagging
is crucial for this analysis. 
The b-tagging algorithm combines several discriminating variables,
as for example, the impact parameter joint probability tag
introduced by ALEPH \cite{aleph},
the $p_{t}$ corrected vertex invariant mass obtained with
the ZVTOP algorithm written for the SLD experiment \cite{zvtop} and a
one-prong charm tag, 
into a feed forward neural network with 12 inputs and 3 output 
nodes, described in Ref. \cite{nn}. 

\begin{figure*}[t]
\centering
\includegraphics[width=65mm]{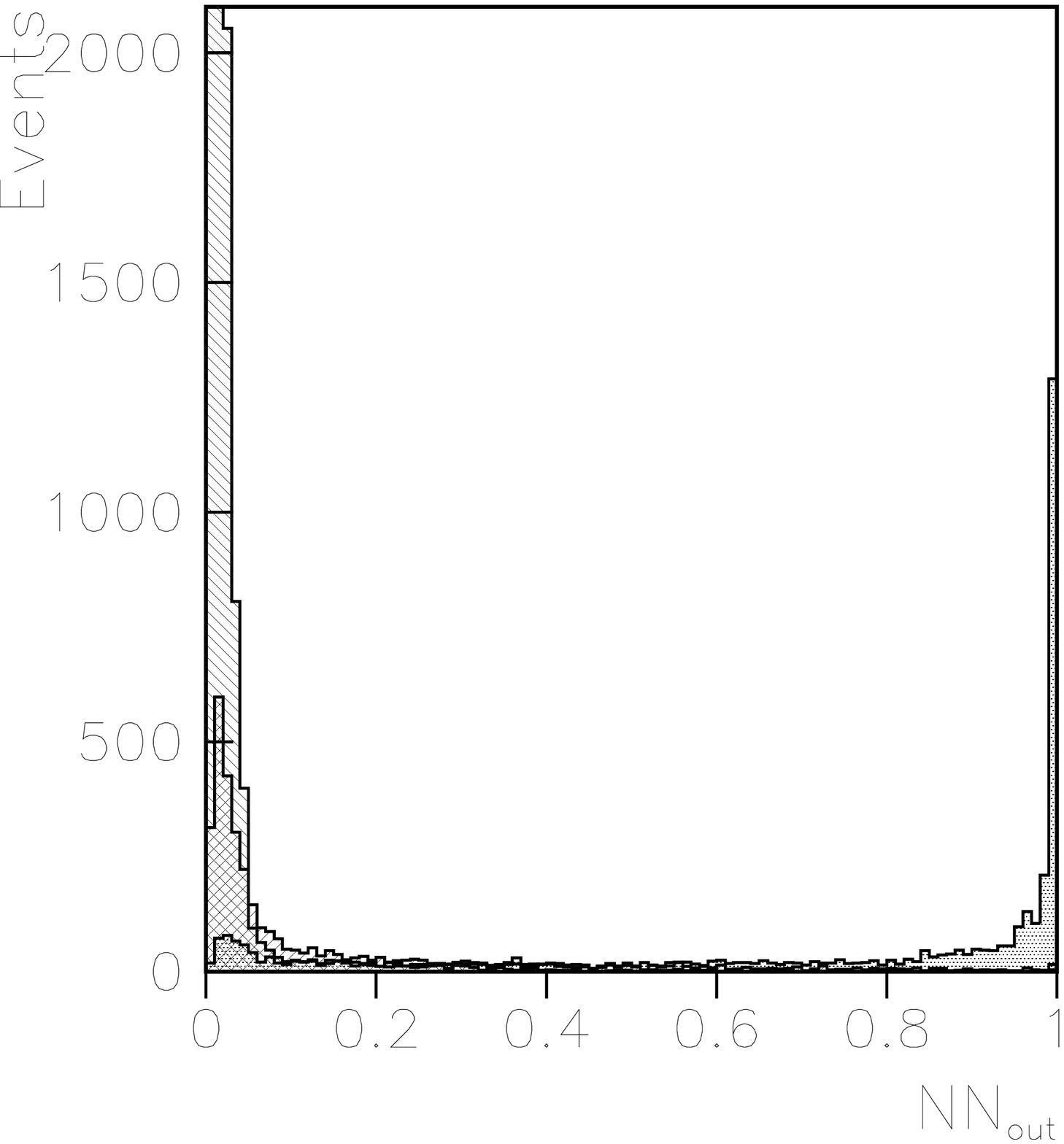}
\includegraphics[width=65mm]{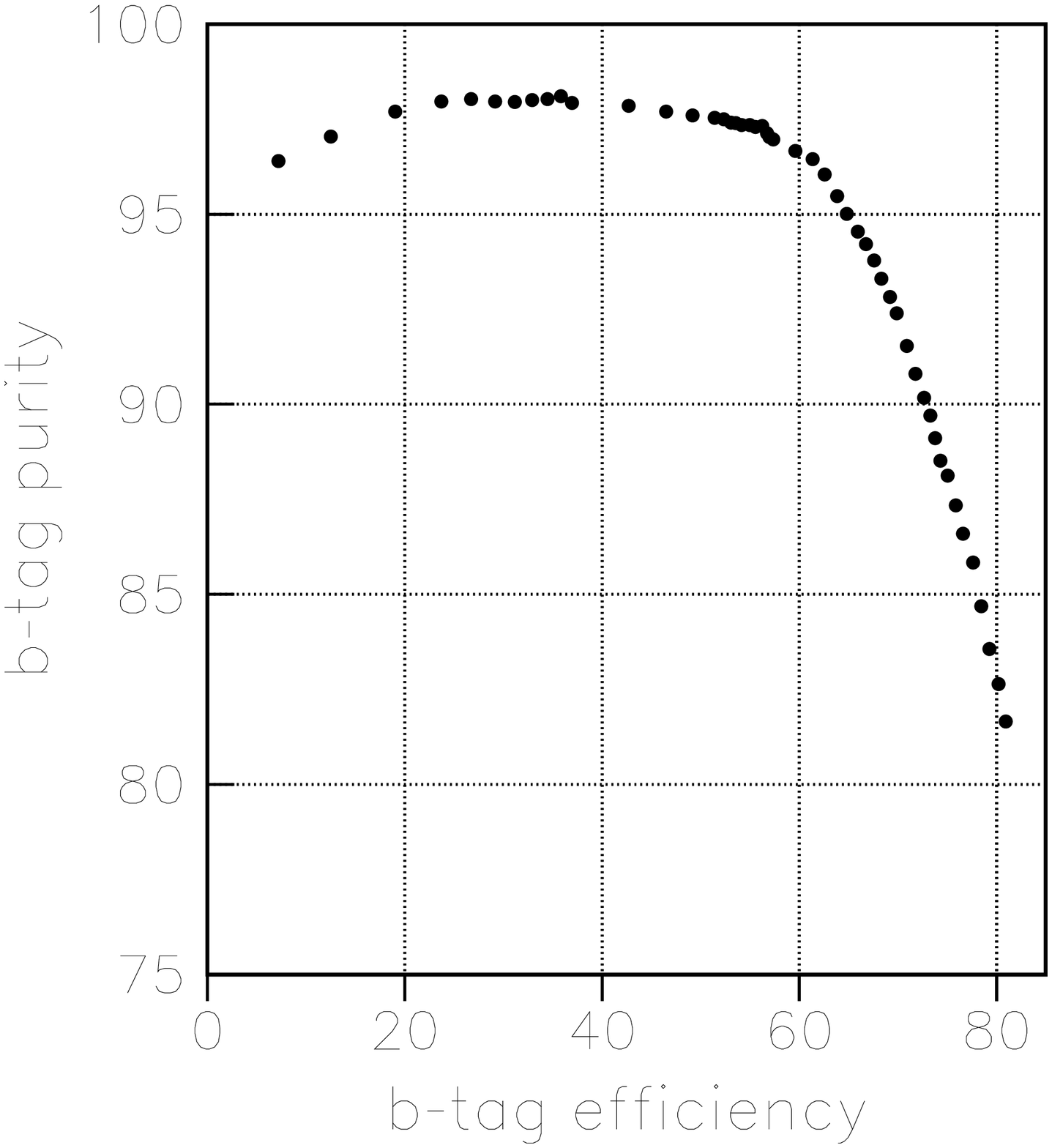} \\
\caption{a) Distribution of the neural network output and
b) efficiency versus purity curve for the neural network based b-tag in 
Z$^{0}$ decays.} 
\label{fig_btag}
\end{figure*}

The distribution of the neural network output $NN_{\rm out}$ to discriminate
b-quark jets
from u-, d-, s- and c-quark jets
is presented in Figure \ref{fig_btag}a. The performance of the neural 
network b-tag in Z$^{0}$
decays is shown in Figure \ref{fig_btag}b. The b-tagging efficiency is 70$\%$ 
and corresponds to a purity of 98$\%$.

The b-quarks coming from the decay of the Higgs boson are highly energetic,
whereas in the case of the background processes
the gluon and one b-quark jet are the most energetic.
In order to reduce the background further we look at the two fastest jets
in the event and require the $NN_{\rm out}$ to be greater than 0.95 for one
jet and greater than 0.2 for the second one.

The expected number of signal and background events are 
summarized in Table \ref{tab_cs}. A total signal efficiency is
estimated to be 36$\%$.

The hadronic cross section for $\gamma \gamma \to$hadrons events,
within the energy range above 2 GeV, is about 400 nb, so that 
about 1 such event is produced per bunch crossing (pileup).
These events obscure the interesting physics processes described in the
previous sections. For this reason this class of events needs to be
included in the PYTHIA simulation for overlap in the next step of this 
analysis.
The HADES \cite{hades} program will be used for this purpose.
A large fraction of this background is distributed at small angles and we
believe that it can be reduced cutting on the polar angle of the tracks  
\cite{aura}. A careful study of this hadronic background is currently
being performed.

\section{Results}

\begin{figure*}[t]
\centering
\includegraphics[width=65mm]{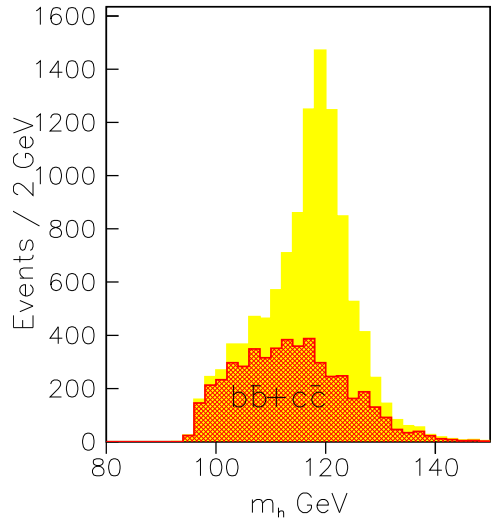}
\caption{Distribution of the reconstructed invariant mass
for the signal and background events.} 
\label{fig_mass}
\end{figure*}

The reconstructed invariant mass for the selected signal and background
events
is shown in Figure \ref{fig_mass}.
To enhance the signal a cut on the invariant mass is
tuned such that the statistical significance of the signal over background
is maximized. Events in the mass region of 114 GeV $ < M_{jj} < $ 126 GeV
are selected. The number of estimated signal and background events
in this window are 4505 and 1698, respectively.

The two photon decay width of the Higgs boson is
proportional to the event
rates of the Higgs signal. The statistical error of the number of signal 
events,
$\sqrt N_{\rm obs}/({\it N} _{\rm obs}-{\it N}_{\rm bkg})$,
corresponds to the statistical error of this measurement. Here $N_{\rm 
obs}$ is the
number of observed events, while $N_{\rm bkg}$ is the number of expected 
background
events.

We obtain
$$\frac{\Delta [\Gamma (\rm h \to \gamma \gamma)\times \rm BR (\rm h \to 
\rm b
\bar{\rm b})]}{[\Gamma (\rm h \to \gamma \gamma) \times \rm BR (\rm h \to 
\rm b
\bar{\rm b})]}=\sqrt N_{\rm obs}/({\it N} _{\rm obs}-{\it N}_{\rm 
bkg})=1.7\% .$$

We conclude that for a Higgs boson with a mass $M_{\rm H}$=120 GeV
we can measure the product $\Gamma (\rm H \to
\gamma \gamma) \times \rm BR (\rm H \to \rm b \bar{\rm b})$
with an accuracy of 1.7$\%$ using an integrated luminosity corresponding 
to one year of data taking at the TESLA Photon Collider.

\end{document}